\documentclass[12pt,letterpaper]{JHEP3}         
\usepackage{epsfig}
\usepackage{cite}

\def\be{\begin{eqnarray}}
\def\ee{\end{eqnarray}}
\newcommand{\nn}{\nonumber}
\newcommand\para{\paragraph{}}

\newcommand{\eqn}[1]{(\ref{#1})}

\newcommand{\spinup}{|\uparrow\,\rangle}
\newcommand{\spindown}{|\downarrow\,\rangle}

\def\Dslash{\,\,{\raise.15ex\hbox{/}\mkern-12mu D}}
\def\Dbarslash{\,\,{\raise.15ex\hbox{/}\mkern-12mu {\bar D}}}
\def\delslash{\,\,{\raise.15ex\hbox{/}\mkern-9mu \partial}}
\def\delbarslash{\,\,{\raise.15ex\hbox{/}\mkern-9mu {\bar\partial}}}
\def\pslash{\,\,{\raise.15ex\hbox{/}\mkern-9mu p}}
\def\calDslash{\,\,{\raise.15ex\hbox{/}\mkern-12mu {\cal D}}}

\newcommand{\Tr}{{\rm Tr}}

\def\lae{\mathrel{\mathop{\smash{\lower .5 ex \hbox{$\stackrel<\sim$}}}}}
\def\lae{\mathrel{\mathop{\smash{\lower .5 ex \hbox{$\stackrel>\sim$}}}}}

\title{Vortices and Impurities}

\author{David Tong and Kenny Wong \\

 Department of Applied Mathematics and Theoretical Physics\\
 University of Cambridge, UK 
\\ {\ } \\
{\tt d.tong, k.wong@damtp.cam.ac.uk}
}

\abstract{We describe the BPS dynamics of vortices in the presence of impurities. We argue that a moduli space of solitons survives the addition of both electric and magnetic impurities. However, dynamics on the moduli space is altered. In the case of electric impurities, the metric remains unchanged but the dynamics is accompanied by a connection term, acting as an effective magnetic field over the moduli space. We give an expression for this connection and compute the vortex-impurity bound states in simple cases. In contrast,  magnetic impurities distort the metric on the moduli space. We show that magnetic impurities can be viewed as  vortices associated to a second, frozen, gauge group. We provide a D-brane description of the dynamics of vortices in product gauge groups and show how one can take the limit such that a subset of the vortices freeze. }

\begin{document}
\pagestyle{plain} \setcounter{page}{1}
\newcounter{bean}
\baselineskip16pt

\section{Introduction and Summary}

Supersymmetric gauge theories have long provided a playground to explore aspects of strongly coupled physics in a controlled manner.  Recently it was shown that one can add charged defects into  field theories in $d=2+1$ dimensions, preserving some amount of  supersymmetry \cite{shamit}.  This offers the prospect of using supersymmetric methods to study strongly coupled phenomena in the presence of doped impurities or lattices\footnote{Supersymmetric defects have already been profitably employed in the context of holography \cite{shamit2,shamit3,shamit4,benicasa,benicasa2}.}.

\para
An important window into the strongly coupled regime is often provided by BPS solitons. (For reviews see, for example, \cite{tasi,misha}). 
Our goal in this paper is to provide a description of the BPS dynamics of vortices in the presence of impurities.  As we will see, the low-energy dynamics is governed, in the familiar geometric fashion, by supersymmetric quantum mechanics on an appropriate moduli space of soliton solutions.

\para
There are two distinct cases: electric and magnetic impurities. In the presence of electric impurities, we show that  the usual BPS vortex solutions remain unchanged. The dynamics of the vortices is described by motion on the moduli space, now augmented by a connection term induced by the impurities.  Using mirror symmetry, this can also be thought of as providing a description for  electrons moving in the background of doped magnetic flux. We study in detail the quantum mechanics of  a single vortex interacting with a delta-function impurity and show how the number of supersymmetric bound states increases with the strength of the defect.

\para
The story is different in the presence of magnetic impurities. Now the vortex solutions themselves are deformed as they approach the defects. Nonetheless, a moduli space of solutions remains, now with a K\"ahler metric which  is distorted by the presence of the impurities. We further show that these magnetic impurities can be thought of as vortices associated to a separate gauge group, frozen in space as they become infinitely heavy.  Finally, we provide a D-brane description of the vortex dynamics, both for vortices in product gauge groups, and for vortices in the presence of magnetic impurities.

\para
Vortices in the presence of electric impurities are described in Section \ref{electricsec}. Magnetic impurities are described in Section \ref{magsec}.
We start, however, in the next section with a review of the basics of vortex dynamics in  theories unspoilt by the presence of dirt.

\section{Vortex Dynamics}

Throughout this paper, we restrict attention to $U(1)$ gauge theories in $d=2+1$ dimensions.   BPS vortices are solutions to the Abelian-Higgs model with a potential that is tuned to lie on the borderline of Type I and Type II superconductivity. In the supersymmetric context, such theories arise naturally with either ${\cal N}=2$ or ${\cal N}=4$ supersymmetry. 

\para
For the purposes of this paper, it will suffice to focus on the electric field $E_i=F_{i0}$, the magnetic field $B=F_{12}$, a scalar $q$ of charge $+1$ and a neutral scalar $\phi$. The action is given by
\be S = - \int d^3x\ \left[ \frac{1}{4e^2}F_{\mu\nu}F^{\mu\nu} +  \frac{1}{2e^2} (\partial_\mu\phi )^2 + |{\cal D}_\mu q|^2 + \frac{e^2}{2}\left(|q|^2-\zeta\right)^2 + \phi^2|q|^2 \right] \label{lag}\ee
where ${\cal D}_\mu q=\partial_\mu q- iA_\mu q$ and $\zeta>0$

\para
The simplest vortex solutions do not involve the neutral scalar $\phi$. Setting the electric field $E_i=0$, we can derive first order vortex equations by the usual Bogomolnyi trick of completing the square in the Hamiltonian, 
\be H &=& \int d^2x\ \frac{1}{2e^2}B^2 + |{\cal D}_iq|^2 + \frac{e^2}{2}\left(|q|^2-\zeta\right)^2\nn\\ 
&=& \int d^2x\ \frac{1}{2e^2}\left(B\mp e^2(|q|^2-\zeta)\right)^2 + |{\cal D}_1q\mp i {\cal D}_2q|^2 \mp B\zeta\nn \\ & \geq&  \mp\int d^2x\ B \zeta = 2\pi |k|\zeta \nn \ee
where $k= - \int B \in {\bf Z}$ is the magnetic flux of the field configuration. We chose to work with the upper-sign, meaning that $k\geq 0$ and we have vortices rather than anti-vortices. The inequality is then saturated if the fields obey the first order vortex equations,
\be B = e^2\left(|q|^2-\zeta\right)\ \ \ \ {\rm and}\ \ \ \ {\cal D}_z q=0\label{vortex}\ee
with $z=x^1+ix^2$. Solutions to these first order equations preserve half the supersymmetries of the theory.

\para
Index theorems show that the general solution to \eqn{vortex} with magnetic charge $k$ has  $2k$ parameters  \cite{erick,taubes}. 
 We write this solution as $ A_i = A_i(x;X^a)$ and $q=q(x;X^a)$ where $X^a$ are the $a=1,\ldots 2k$ collective coordinates. These can be thought of as the positions of $k$ vortices. They  parameterise the vortex moduli space,
\be {\cal M}_k \cong {\bf R}^2\times \tilde{\cal M}_k\label{factor}\ee
Here ${\bf R}^2$ describes the centre of mass of the vortex system, while $\tilde{\cal M}_k$ parameterises the relative separations.

\subsection{Motion on the Moduli Space}\label{motionsec}

The low-energy dynamics of the vortices arises by allowing the collective coordinates to depend on time, $X^a=X^a(t)$, resulting in a sigma-model with target space  ${\cal M}_k$ \cite{manton}. Crucially, in order to derive the relevant metric on ${\cal M}_k$,  some attention must be paid to gauge fixing. Since this aspect of the story will prove to be important once we introduce the electric impurities, we spend some time describing it in detail here.

\para
At issue is the Gauss law constraint for the gauge field $E_i=\partial_iA_0-\partial_0A_i$, which reads
\be \frac{1}{e^2}\partial_i E_i =iq^\dagger {\cal D}_0q - iq {\cal D}_0q^\dagger\label{gauss}\ee
Naively promoting $X^a\rightarrow X^a(t)$ will not, in general, lead to a solution of Gauss' law; we need to also turn on $A_0$. This can be accomplished as follows:  for each collective coordinate $X^a$, one constructs a zero mode that obeys the linearised vortex equations,
\be \delta_a A_i = \frac{\partial A_i}{\partial X^a} +\partial_i\alpha_a\ \ \ \ ,\ \ \ \ \  \delta_aq = \frac{\partial q}{\partial X^a} + i\alpha_a q\nn\ee
In each case, the first term comes from differentiating the field with respect to the collective coordinate, and the second term is a  gauge transformation. The functions $\alpha_a(x;X)$ are chosen to meet the gauge fixing condition inherited from Gauss' law \eqn{gauss},
\be -\frac{1}{e^2}\partial_i\,\delta_a A_i = iq^\dagger\delta_a q - i q\,\delta_aq^\dagger\label{zeromode}\ee
Setting  $A_0 = - \alpha_a \dot{X}^a$, one finds that Gauss' law \eqn{gauss} is obeyed, with  the covariant derivatives of the field are related to the zero modes,
\be E_i=-\delta_a A_i\,\dot{X}^a \ \ \ \ ,\ \ \ \ \  {\cal D}_0 q= \delta_a q\,\dot{X}^a \nn\ee
Substituting these expressions into the original action \eqn{lag} gives the promised geometrical description of the vortex dynamics in terms of motion on the moduli space,
\be S_{\rm vortex} = \int dt\ \frac{1}{2}g_{ab}(X) \dot{X}^a\dot{X}^b\nn\ee
where the metric on ${\cal M}_k$ is given by the overlap of zero modes,
\be g_{ab} = \int d^2x\ \left(\frac{1}{e^2}\delta_aA_i\,\delta_bA_i + \delta_aq\,\delta_bq^\dagger + \delta_aq^\dagger\delta_bq\right)\label{metric}\ee
It is not difficult to show that this metric is smooth and K\"ahler. Indeed, the latter property follows from the requirements of supersymmetry.  In $d=2+1$ dimensional theories with ${\cal N}=4$ supersymmetry, the vortex dynamics preserves ${\cal N}=(2,2)$; in theories with ${\cal N}=2$ supersymmetry, the vortex dynamics preserves ${\cal N}=(0,2)$. Both require a K\"ahler target space\footnote{This ${\cal N}=(p,q)$ characterisation of supercharges in quantum mechanics is inherited from $d=1+1$ dimensions. After dimensional reduction to $d=0+1$, the ${\cal N}=(0,2)$ algebra has a $U(1)_R$ R-symmetry, while the ${\cal N}=(2,2)$ algebra has an $SU(2)_R\times U(1)_R$ R-symmetry.}.

\subsubsection*{The Connection}

As we have reviewed above, the vortex dynamics induces a natural metric $g_{ab}$ on the moduli space. However, equally important for the purpose of the present paper are the functions $\alpha_a(x;X)$ which were introduced to ensure that Gauss law is obeyed. These can be thought of as a $U(1)$ connection over ${\cal M}_k$. 

\para
It will prove useful to give an example of this connection. We look at the simplest case of the two collective coordinates corresponding to the centre of mass motion of the vortices. Changing notation slightly, we use $X$ to denote the position of the centre of mass, with $\tilde{X}$ parameterising the relative moduli space $\tilde{\cal M}_k$. We write the general solution as $A_i=A_i(x-X;\tilde{X})$ and $q=q(x-X;\tilde{X})$.  The translational zero modes associated to the centre of mass motion are
\be \delta_j A_i = -\partial_j A_i +\partial_i\alpha_j\ \ \ \ ,\ \ \ \ \ \delta_jq = -\partial_j q +i\alpha_jq\label{zero}\ee
where $\alpha_j$ is determined by the gauge fixing  condition \eqn{zeromode}. It is simple to check that, in this case, the connection $\alpha_j$ is nothing other than the background gauge field itself,
\be \alpha_j (x-X; \tilde X)= A_j (x-X; \tilde X) \label{connection}\ee
One immediate consequence of this is that the translational zero modes can be written in the simple form $\delta_jA_i = -B\epsilon_{ij}$ and $\delta_j q = -{\cal D}_jq$. In the next section we will see how this connection plays a more direct  role in the vortex dynamics. 

\section{Electric Impurities}\label{electricsec}

In this section, we dope our theory with a static, electric charge density $\rho(x)$. This is achieved by adding a source term $\rho(x)A_0$ for the gauge field (which is gauge invariant provided that $\dot{\rho}=0$). It was shown in  \cite{shamit} that such a perturbation preserves half the supersymmetry, provided that it is accompanied by a similar source for the neutral scalar $\phi$. We therefore sully the action \eqn{lag} by adding the impurity
\be S_{\rm impurity} = \int d^3x \ \rho(x)(A_0-\phi)\label{eimp} \ee
Now neither the electric field nor $\phi$ are passive bystanders. In vacuum, both are sourced by the impurities. Indeed, both obey the same equation
\be \frac{1}{e^2}\partial_iE_i= 2|q|^2A_0 +\rho(x)\ \ \ \ ,\ \ \ \ \frac{1}{e^2}\partial^2\phi = 2|q|^2\phi + \rho(x)\label{easy}\ee
Because the equations are the same,  the solutions are the same and $A_0=\phi$. This has consequence for other fields. In particular, it  means that there is no knock-on effect on the field $q$ which can happily remain in its vacuum value $|q|^2=\zeta$ when solving \eqn{easy}.

\para
Note that if $\zeta=0$  then both $A_\mu$ and $\phi$ are massless and suffer infra-red logarithmic divergences unless $\int \rho(x) = 0$. In what follows, we will be interested in the case with $\zeta\neq 0$ where there is no such restriction. 

\subsection{Vortex Dynamics}

We would like to ask: what becomes of vortices in the presence of these electric impurities? We start by focussing on the static solutions, before turning to their dynamics. 

\para
In fact, it is simple to check that the static vortex solutions remain unchanged in the presence of the source $\rho(x)$ as long as $A_0=\phi$. This means that   solutions to the first order vortex equations \eqn{vortex} continue to solve the full second order equations of motion in the theory with impurities. However,  the presence of the vortices does feed back onto the electric field and  the profile of $\phi$, both of which now solve \eqn{easy} with $|q|^2$ given by the vortex profile rather than its vacuum value.  

\para
The upshot of this simple discussion is that the moduli space of vortices in the presence of electric impurities is again given by ${\cal M}_k$.  But how do the vortices move?

\para
We again implement the moduli space approximation, promoting the collective coordinates to dynamical degrees of freedom: $X^a\rightarrow X^a(t)$. And, again, the zero modes $\delta_aA_i$ and $\delta_a q$ include a compensating gauge transformation as in \eqn{zero}. The linearised vortex equations are unchanged. Meanwhile, Gauss' law reads
\be \frac{1}{e^2}\partial_i E_i = iq^\dagger {\cal D}_0q - iq{\cal D}_0q^\dagger + \rho(x)\nn\ee
and is solved by setting
\be
A_0=\phi -\alpha_a\dot{X}^a\nn\ee
 where $\phi$ is the solution to the static equations \eqn{easy} and the gauge connection $\alpha_a$ is determined by solving the same equation \eqn{zeromode} that we had in the absence of impurities. This is important: it means that not only are the static vortex solutions \eqn{vortex} unchanged by the presence of electric impurities, but the zero modes \eqn{zero} are also unchanged. 

\para
Our ansatz for the time-dependent fields is now
\be E_i=\partial_i\phi -\delta_aA_i\,\dot{X}^a\ \ \ \ ,\ \ \ \ \ {\cal D}_0q = -i\phi q + \delta_aq\,\dot{X}^a\label{moving}\ee
With this in hand, we are now almost ready to substitute the time dependent fields into the action. There is one remaining subtlety: we work to leading order in the charge density and ignore the term $\dot{\phi}^2$ in the action. (The same approximation is required in the study of Chern-Simons vortex dynamics \cite{kimyeong,me} although, in that case, one can show that the end result is actually exact. Indeed, naively it appears that the kinetic terms arising from $\dot{\phi}^2$ do not preserve the K\"ahler property of the target space, strongly suggesting that these terms would not contribute in an exact treatment here either). As before, the $B^2$ term and the $|{\cal D}_iq|^2$ combine to give the vortex mass. The remaining terms are
\be S_{\rm vortex} &=& \int d^3x\  \left[ \frac{1}{2e^2}E_i^2 + |{\cal D}_0q|^2  -\frac{1}{2e^2}\partial_i\phi^2-\phi^2|q|^2 -  \rho(x)(A_0-\phi) \right] \nn\ee
Substituting the background fields \eqn{moving}, we find an expression for the vortex dynamics
\be
S_{\rm vortex} 
= \int dt\ \frac{1}{2}g_{ab}(X)\dot{X}^a\dot{X}^b + {\cal A}_a(X)\dot{X}^a  \label{evortex}\ee
Here $g_{ab}$ is the usual metric \eqn{metric} on the vortex moduli space. But we see that the dynamics is now augmented by a connection term, ${\cal A}_a$, induced by the impurities. We refer to ${\cal A}_a$ as the {\it dirty connection}. It plays the role of an effective magnetic field over the vortex moduli space. After an integration by parts, the dirty connection is given by
\be {\cal A}_a(X) = \int d^2x\ \phi\left( \frac{1}{e^2}\partial_i\,\delta_aA_i - iq\,\delta_aq^\dagger +iq^\dagger\delta_aq\right)+ \rho(x)\alpha_a \nn\ee
But the term in brackets vanishes courtesy of the gauge fixing condition on the zero modes \eqn{zeromode}. We're left with the simple expression for the dirty connection ${\cal A}_a$ over the moduli space in terms of the compensating gauge connection $\alpha_a$,
\be {\cal A}_a(X) = \int d^2x\ \rho(x)\alpha_a(x,X)\label{a}\ee
It is pleasing to see the abstract connection $\alpha_a$ promoted to play a physical role in the vortex dynamics. 

\subsubsection*{An Example}

We can illustrate the role of the dirty connection through a simple example. Consider a single $k=1$ vortex of mass $M=2\pi\zeta$, moving in the presence of a delta-function electric impurity,
\be \rho(x) = g \delta(x)\label{delta}\ee
where $g$ is dimensionless and we require $g e^2/\zeta \ll 1$ for the validity of our approximation.
The vortex has only two translational zero modes. As we saw in \eqn{connection}, the corresponding gauge transformation is nothing but the background gauge field of the vortex: $\alpha_i = A_i$. In this case, we see that the dynamics of the vortex is just
\be L_{\rm 1-vortex} = \frac{1}{2}M\dot{X}^2 + gA_i(X)\dot{X}^i\label{onevortex}\ee
This, of course, is to be expected. The extra term is the Lorentz force law, now due to a localised electric flux acting on a magnetic particle. If the vortex encircles the impurity, its wavefunction  picks up an Aharonov-Bohm phase given by $g\oint A\cdot dX = g\int B = 2\pi g$.

\subsection{Supersymmetric Vortex Dynamics}

We have seen that the low-energy dynamics of bosonic vortices is governed by the effective action \eqn{evortex}. We would now like to understand the role played by fermionic zero modes. 

\para
Let us first review what happens in the absence of impurities. Vortices in the ${\cal N} = 2$ theory are $1/2$-BPS. They preserve two real supercharges  and their low-energy dynamics is described by an ${\cal N} = (0,2)$ sigma-model. In the theory with ${\cal N}=4$ supersymmetry, the vortices are also $1/2$-BPS, now preserving four real supercharges.  One way to see this is to note that the three-dimensional  theory enjoys an $SU(2)_N$ R-symmetry. The vortices  inherit one pair of supercharges from the ${\cal N}=2$ theory;  acting on this pair with  the R-symmetry gives rise to two further, linearly independent,  supercharges that are also preserved in the background of the vortices. This ensures that the vortex dynamics is now governed by a sigma-model with ${\cal N}=(2,2)$ supersymmetry.

\para
Adding electric impurities changes this. Vortices in the ${\cal N} = 2$ theory remain $1/2$-BPS, preserving two real supercharges. (It is simple to check that the Bogomolnyi equations are compatible with the requirement  $A_0 = \phi$ imposed by the impurity). However, in the ${\cal N}=4$ theory the electric impurities break the $SU(2)_N$ R-symmetry. This is because the chosen scalar field $\phi$ is now part of an $SU(2)_N$ triplet. With no $SU(2)_N$ symmetry, there is no further enhancement of the number of supercharges. The result is that, when electric impurities are present, vortices in the ${\cal N}=4$ theory preserve just two real supercharges. 
This discussion means that the dirty connection  in the sigma-model \eqn{evortex} has an ${\cal N} = (0,2)$ supersymmetric completion, and this is true for both
the ${\cal N} = 2$ and the ${\cal N} = 4$ theories.

\para
It is simple to write down a such a connection  term using ${\cal N}=(0,2)$ chiral superfields, $Z^a$. (We're indulging in a slight abuse of notation here, with $a$ now labelling complex coordinates on the target space rather than real coordinates). Each superfield houses a complex scalar $z^a$ and a single complex Grassmann object $\psi^a$. We introduce a real function $C(Z^a,Z^{a\dagger})$. Integrated over all of superspace, we have 
\be \int d^2\theta\, C(Z^a, Z^{a\dagger}) =  i\partial_aC\,\dot{z}^a -i\bar{\partial}_a C \,\dot{z}^{a\dagger} + 2\bar{\partial}_a\partial_b C\,\bar{\psi}^a\psi^b\label{super}\ee
We see that the holomorphic part of the connection is given by ${\cal A}_a = i\partial_aC$.

\subsubsection*{Quantum Mechanics of a Single Vortex}

We now focus on a single vortex moving in the background of a delta-function impurity \eqn{delta}.  Our goal is to count the number of supersymmetric bound states between the vortex and impurity. We will show that the number of such states is determined by the integer part of $g$,  the (dimensionless) strength of the impurity.

\para 
The vortex has a single bosonic degree of freedom, $z=X^1+iX^2$ and a single Grassmann degree of freedom $\psi$. The ${\cal N}=(0,2)$ supersymmetric completion of the low-energy dynamics \eqn{onevortex} is
\be L_{\rm 1-vortex} = \frac{1}{2}M |\dot{z}|^2 + \frac i 2 M\bar{\psi}\dot\psi +g A_z\dot{z} + gA_{\bar{z}} \dot{z}^\dagger - \frac 1 2 gB \bar{\psi}\psi\label{susyact}\ee
where both $A_z$ and $B=-2iF_{z\bar{z}}$ are the fields of the vortex profile. 

\para
To show that the connection term can arise from the ${\cal N}=(0,2)$ superspace integral \eqn{super}, we need to find a real function $C$ such that $dC=g(A_1dx^2 - A_2 dx^1)$. Since this function $C$ will also play a role in determining the supersymmetric ground state wavefunctions, we take some time to explain how to construct it.  We work in a gauge such that $\partial_i A_i=0$. For a single vortex, this can be achieved by  the ansatz,
\be A_1 = -\frac{x^2}{2r}a(r)\ \ \ \ ,\ \ \ \ A_2 = +\frac{x^1}{2r}a(r)\label{goinground}\ee
Then we can construct $C(r)$ by integrating $g(A_1dx^2-A_2dx^1)$ from the origin to the point $r$. Any path will do because, by construction, $g(A_1dx^2-A_2dx^1)$ is closed and the complex plane has trivial cohomology. This allows us to write
\be C(r) = -\frac{g}{2}\int_0^r dr'\ a(r')\nn\ee
Let us now turn to the quantization of the theory. As usual, quantizing the fermions splits the Hilbert space into two sectors, which can be thought of as spin up $\spinup$ and spin down $\spindown$. These obey
\be \psi\spindown = \bar{\psi}\spinup =0, \qquad \psi\spinup = \sqrt{\frac 2 M} \spindown, \qquad \bar{\psi} \spindown = \sqrt{ \frac 2 M} \spinup \nn\ee
The most general    wavefunction then takes the form
\be |\omega\rangle = f(z,\bar{z})\spinup + h(z,\bar{z}) \spindown\nn\ee
The ground state wavefunctions have vanishing energy, $H\,|\omega\rangle =0$, where the Hamiltonian can be written, as usual, in terms of the supercharges $Q$ by  $H = \{Q, Q^\dagger\} $. It follows by a standard argument that the ground state wavefunctions are precisely those which are annihilated by the supercharges 
\be Q = i(p - A_z)\psi\ \ \ ,\ \ \ Q^\dagger= -i(p^\dagger  - A_{\bar{z}}) \bar{\psi}\nn\ee
where the canonical momentum is given by $p= \frac 1 2 M \dot{z}^\dagger + A_z$. We are therefore looking for wavefunctions that obey
\be Q\,|\omega\rangle = Q^\dagger \,|\omega\rangle =0\nn\ee
This condition can be written as 
\be Q\,|\omega\rangle = e^{-C}\partial_z(e^Cf)\,\spindown =0 \ \ \ \ ,\ \ \ \ Q^\dagger\,|\omega\rangle = - e^{+C}\partial_{\bar{z}}(e^{-C}h)\, \spinup =0 \nn\ee
Solutions to these equations are straightforward. They are given by  $f=e^{-C}\tilde{f}(\bar{z})$ and $h=e^{+C}\tilde{h}(z)$ where $\tilde{f}(\bar{z})$ is any anti-holomorphic function and $\tilde{h}(z)$ is any holomorphic function. However, not all of these are admissible wavefunctions: we also require that they are $L^2$-normalizable. To see the implications of this requirement, we need to look more closely at the asymptotic behaviour of $C(r)$. 

\para
It is simple to check that the asymptotic fall-off of the vortex gauge field $a(r)$ defined in \eqn{goinground} is given by $a(r)\sim 2/r$ (up to exponentially small corrections). This fall-off is necessary to cancel the divergent gradient term that arises from the winding of $q$ and, in turn, gives rise to the quantised magnetic flux $\int B=-2\pi$. This means that, at large $r$, the asymptotic behaviour of $C$ is given by
\be C(r) \sim - g\log r\nn\ee
We then have the following normalizable supersymmetric ground states.
\begin{itemize}
\item \underline{$g>0$:} \ The ground states take the form
\be |\omega\rangle = e^Cz^n\,\spinup\ \ \ \ \ \ {\rm with}\ \  n=0,1,\ldots,< |g|-1\nn\ee
To illustrate what the notation means, it's easiest to go through some examples. There are no ground states for $0<g\leq 1$. In particular, when $g=1$, the $n=0$ wavefunction has logarithmically divergent norm and does not provide a good ground state. For $1<g\leq 2$ there is a single ground state; for $2<g\leq 3$ there are two ground states, and so on.
\item \underline{$g<0$:}\  \ For $g<0$, the story is the same, but with the ground states lying in the spin down sector,
\be |\omega\rangle = e^{-C}\bar{z}^n\,\spindown\ \ \ \ \ \ {\rm with}\ \  n=0,1,\ldots,< |g|-1\nn\ee
Again, there are no ground states for $-1 \leq g <0$; the first ground state appears when $g<-1$. 
\end{itemize}
The counting above is (almost) what one would get by naive semi-classical analysis. It's well known that spin 1/2 particles of electric charge $g$ have zero energy in a constant magnetic field, with density of states $gB/2\pi$. This agrees with the above analysis, modulo the question of logarithmic divergences for states when $g\in {\bf Z}$. 

\para
Note that for $g\notin{\bf Z}$, the impurity-vortex bound states act as abelian anyons.

\section{Magnetic Impurities}\label{magsec}

We now turn to the effect of magnetic impurities on vortices. The impurities are comprised of a fixed, static source term $\sigma(x)$ for the magnetic field,
\be S_{\rm impurity} = -\int d^3x\ \sigma(x)B\label{mimp}\ee
It was shown in \cite{shamit} that such an impurity preserves half of the supersymmetry if the auxiliary $D$ field is similarly sourced. After integrating out this $D$ term, we're left with the action
\be S = - \int d^3x\ \left[ \frac{1}{4e^2}F_{\mu\nu}F^{\mu\nu} +|{\cal D}_\mu q|^2 +  \frac{e^2}{2}\left(|q|^2-\zeta-\sigma(x)\right)^2 + \sigma(x)B \right] \label{lag2}\ee
where we have omitted the neutral scalar $\phi$ because it will play no further role in our discussion. The purpose of this section is to describe the dynamics of vortices in this theory. 

\para
As shown in \cite{shamit}, the theory with impurities admits first order vortex equations. These can be derived using the standard Bogomolnyi trick described in Section 2; they are given by
\be B = e^2\left(|q|^2 - \zeta-\sigma(x)\right)\ \ \ \ ,\ \ \ \ \ {\cal D}_zq=0\label{newvortex}\ee
Fields obeying these equations solve the full second order equations of motion and describe an object with mass $M= 2\pi k \zeta$, where $k=-\int B\in {\bf Z}^+$ is the winding number.

\para
One can ask whether there are solutions to \eqn{newvortex} and, if so, how many?  Energetic considerations suggest that the vortices feel neither an attractive nor repulsive force towards the defect: the seeming change in their mass due to the magnetic source \eqn{mimp} is exactly compensated by the varying scalar expectation value of $\zeta+ \sigma(x)$. It seems at least plausible therefore that a full $2k$-dimensional moduli space exists. 

\para
While we have not done a full analysis along the lines of \cite{taubes}, there are a number of further arguments, all of which point to the existence of a $2k$-dimensional moduli space of solutions.  
First, the usual index theorem \cite{erick} goes through without hitch in the presence of impurities. The reason is simple: the index theorem counts the number of solutions to the linearised vortex equations. But the linearised versions of \eqn{newvortex} do not depend on the source $\sigma(x)$. This means that if there is a single solution to \eqn{newvortex} then there are $2k$ linearised deformations. In particular,  for a rotationally symmetric source $\sigma(x)$, it is simple to check that there must exist a rotationally symmetric solution to \eqn{newvortex} centered on the impurity, together with the corresponding $2k$ linearised deformations. (Admittedly, this is not quite enough to guarantee a full moduli space even in this case).  

\para
Further evidence comes through more indirect means.  We will shortly relate the existence of solutions to \eqn{newvortex} to the existence of vortices in product gauge groups where a full moduli space of solutions does indeed exist. Finally,  the D-brane picture (to be developed in the next section) also shows the existence of $2k$ independent objects. 

\para
For now, we assume that \eqn{newvortex} admit a $2k$-dimensional moduli space of solutions which we denote as ${\cal M}_k(\sigma)$. This means that the usual moduli space approximation, summarised in Section \ref{motionsec}, can be employed, and the low-energy vortex dynamics is once again described as motion on ${\cal M}_k(\sigma)$, with a metric defined in terms of the overlap of zero modes as in \eqn{metric}. However, because the solutions are deformed as the vortices approach the defects, the metric on ${\cal M}_k(\sigma)$ depends on the function $\sigma(x)$. In particular,  since we no longer have translational invariance, the moduli space no longer factorizes as \eqn{factor}.

\para
Clearly it would be of interest to get a better handle on the metric over ${\cal M}_k(\sigma)$. This seems hard; even in the absence of impurities, the metric on the two-vortex moduli space is not known analytically. Here we instead show that there is a different way of viewing the magnetic impurities that sheds light on the problem. 

\subsection{Impurities as Frozen Vortices}

The purpose of this section is to show how the magnetic impurities in the action \eqn{lag2} can be viewed as heavy, frozen vortices that sit in a different gauge group. 

\para
 To this end, we consider a theory with product gauge group $\hat{U}(1)\times \tilde{U}(1)$. We introduce two, charged scalar fields: $q$ carries charge $(+1,-1)$ and $p$ carries charge $(0,+1)$. The action is
\be S =- \int d^3x && \left[ \frac{1}{4e^2}\hat{F}_{\mu\nu}\hat{F}^{\mu\nu} + \frac{1}{4\tilde{e}^2}\tilde{F}_{\mu\nu}\tilde{F}^{\mu\nu} + |{\cal D}_\mu q|^2 + |{\cal D}_\mu p|^2 \right. \nn\\  &&  \ \ \ \  \left. +\ \frac{e^2}{2} \left(|q|^2-\zeta\right)^2 + \frac{\tilde{e}^2}{2}(-|q|^2+|p|^2-\tilde{\zeta})^2 \right]  \label{product}\ee
Here ${\cal D} q =\partial q -i\hat{A} q + i\tilde{A}q$ and ${\cal D}p = \partial p -i\tilde{A}p$. Our goal is to show how we can take a particular limit so that the translationally invariant theory \eqn{product} reduces to the impurity theory \eqn{lag2}. 

\para
In the following, we assume that $\zeta > 0$ and $\tilde{\zeta} > -\zeta$.
The vacuum of the theory is $|p|^2 = \zeta+\tilde{\zeta}$ and $|q|^2=\zeta$ and both $U(1)$ factors of the gauge group are spontaneously broken. The theory has two types of vortices, one for each gauge group. The static vortex equations can be derived using the now-familiar Bogomolnyi  trick. They read
\be \hat{B} = e^2(|q|^2 -\zeta) \ \ \ ,\ \ \ {\cal D}_zq=0\label{vib1}\ee
and
\be \tilde{B} = \tilde{e}^2(-|q|^2+|p|^2-\tilde{\zeta})\ \ \ ,\ \ \ {\cal D}_zp=0\label{vib2}\ee
%
Solutions to these equations have energy
\be M = -\int d^2x\ \hat{B}\zeta +\tilde{B}\tilde{\zeta} = 2\pi \hat{k} \zeta+ 2\pi \tilde{k}\tilde{\zeta} \nn\ee
where $\hat{k}$ and $\tilde{k}$ are magnetic fluxes for $\hat{U}(1)$ and $\tilde{U}(1)$ respectively. However, it will turn out to be somewhat more revealing to write the mass of the vortex in terms of the winding numbers of the two scalar fields. The winding number of $p$ is simply $\tilde{n}=\tilde{k}$. But, because the field $q$ is charged under both $\hat{U}(1)\times \tilde{U}(1)$, its winding number is $n=\hat{k}-\tilde{k}$. The mass is then 
\be M = 2\pi n\zeta + 2\pi \tilde{n}(\tilde{\zeta}+\zeta)\label{nn}\ee
This way of writing the mass is more useful for two reasons. First, solutions to the vortex equations \eqn{vib1} and \eqn{vib2} only exist if $n,\tilde{n}\geq 0$. Secondly, it is the winding of the scalar fields that determines the identities of the vortices. In particular, when the vortices are well separated, they behave like $n$ objects with mass $\zeta$ and $\tilde{n}$ objects with mass $\tilde{\zeta}+\zeta$. (This will be seen clearly in the D-brane picture that we develop in the next section). 

\para
Now we can explain how one can freeze vortices in this model. We send $\tilde{\zeta}\rightarrow \infty$, so that the $\tilde{n}$ vortices become very heavy. Physically, we expect that the $n$ light vortices will move in the background of the $\tilde{n}$ heavy ones. Our goal is  to implement this physical expectation mathematically. 

\para
As we take $\tilde{\zeta}\rightarrow \infty$, it's not just the $\tilde{n}$ vortices that become heavy. From the original action \eqn{product}, we see that the elementary fields $\tilde{A}$ and $p$ are also heavy. However, in relativistic field theories, very massive fields do not freeze; quite the contrary, they are the fast moving degrees of freedom which respond quickly to what's going on around them. This is precisely why we can integrate them out and forget about them.

\para
In contrast, in non-relativistic physics, heavy particles are the slow-moving degrees of freedom (e.g. nuclei in the Born-Oppenheimer approximation). These act very much like frozen impurities. 

\para
Now, in our product gauge theory, as $\tilde{\zeta}\rightarrow \infty$, the fields $\tilde{A}$ and $p$ play the roles of both fast and slow degrees of freedom! If we fix the asymptotic winding $\tilde{n}$ of $p$, then the vortex degrees of freedom are slow-moving particles, while all other fluctuations of of $\tilde{A}$ and $p$ are fast moving.  To implement this, we require that the fields $\tilde{A}$ and $p$ obey the vortex equations \eqn{vib2}. Note that this doesn't fix them completely. We will view $\tilde{B}$ as fixed, but if $q$ fluctuates then $p$ responds so that \eqn{vib2} still holds.

\para
Meanwhile, $q$ is charged under both gauge groups, with covariant derivative
\be {\cal D}_\mu q = \partial_\mu q - i\hat{A}_\mu q + i\tilde{A}_\mu q\nn\ee
We have already fixed $\tilde{A}$ to carry magnetic flux $\tilde{n}$. But we do not yet wish to predetermine the winding of $q$. For this reason, we write 
\be \hat{A}_\mu = \tilde{A}_\mu + A_\mu \nn\ee
Then ${\cal D}q = \partial q - iAq$ and the gauge field $A$ carries magnetic flux $n$, equal to the winding of $q$. We now substitute these fixed relationships back into the Hamiltonian. Completing the square for $\tilde{B}$ and ${\cal D}p$, we obtain
\be H &=& \frac{1}{2e^2} \hat{B}^2 + |{\cal D}_iq|^2 + \frac{e^2}{2}(|q|^2 - \zeta)^2  -\tilde{B}|q|^2 -\tilde{B}\tilde{\zeta} \nn \\ && + \ \frac 1 {2\tilde e^2} \left(\tilde B - \tilde e^2(-|q|^2 + |p|^2 - \tilde \zeta ) \right)^2 + |{\cal D}_1 q - i {\cal D}_2 q|^2\nn\ee

\noindent
The vortex equations \eqn{vib2} set the second line to zero, and the remaining Hamiltonian for $\hat{A}$ and $q$ reads
\be H &=& \frac{1}{2e^2} \hat{B}^2 + |{\cal D}_iq|^2 + \frac{e^2}{2}(|q|^2 - \zeta)^2  -\tilde{B}|q|^2 -\tilde{B}\tilde{\zeta}\nn \\ &=& \frac{1}{2e^2} B^2 + |{\cal D}_iq|^2 + \frac{e^2}{2}(|q|^2 - \zeta)^2  -\tilde{B}|q|^2 +\frac{1}{e^2}B\tilde{B} + \frac{1}{2e^2}\tilde{B}^2-\tilde{B}\tilde{\zeta} \nn\\ 
&=&  \frac{1}{2e^2} B^2 + |{\cal D}_iq|^2 + \frac{e^2}{2}(|q|^2 - \zeta -\sigma(x) )^2 + \sigma(x)B - \tilde{B}(\zeta+\tilde{\zeta}) \ee
The final term is just the mass of the frozen vortices. The remaining terms are precisely the Hamiltonian for the impurity theory \eqn{lag2}, with the magnetic source given by $\sigma(x) = \tilde{B}/\tilde{e}^2$.

\para
The upshot of this analysis is that the dynamics of vortices in the presence of impurities is a limiting case of the dynamics of vortices in product gauge groups. 

\subsection{D-Branes}\label{dbranesec}

In this final section, we provide a description of the dynamics of vortices using D-branes. For vortices in $U(N)$ gauge groups, such a description was given in \cite{vib}. Here we extend this to product gauge groups.

\begin{figure}[htb]
\begin{center}
\epsfxsize=5.9in\leavevmode\epsfbox{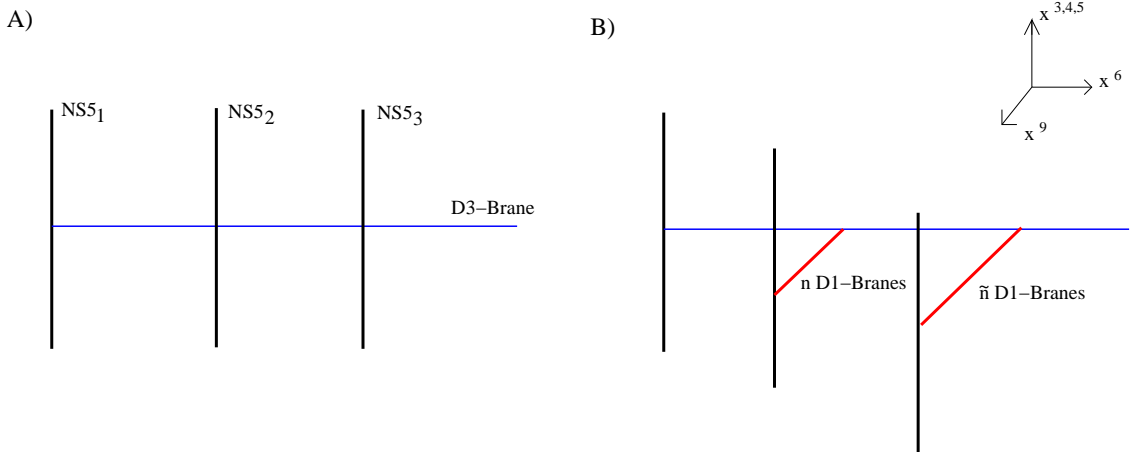}
\end{center}
\caption{The Hanany-Witten set-up for the $d=2+1$ dimensional gauge theory. D1-branes, in red, are vortices in the Higgs phase.}
\label{fig1}
\end{figure}

\para
For concreteness, we focus exclusively on the $\hat{U}(1)\times \tilde{U}(1)$ gauge group described in the previous section (although the results generalise in an obvious manner to higher rank groups and larger linear quivers). The D-brane configuration  consists of three NS5-branes extended in the $012345$ directions, with a single D3-brane spanning the $0126$ directions \cite{hw}. This configuration is shown in Figure \ref{fig1}.

\para
The low-energy effective theory for the D3-branes is given by the ${\cal N}=4$ supersymmetric completion of the $d = 2+1$ dimensional theory \eqn{product}. The parameters of this low-energy theory are determined by the positions of the NS5-branes. Ignoring factors of $g_s$ and $\alpha'$ for simplicity (see, e.g. \cite{vib}, for a more careful treatment), the gauge coupling constants are determined by the positions of the NS5-branes in the $X^6$ direction, 
\be \frac{1}{e^2} = \left.X^6\right|_{NS5_2} - \left. X^6\right|_{NS5_1}\ \ \ ,\ \ \  \frac{1}{\tilde{e}^2} = \left.X^6\right|_{NS5_3} - \left. X^6\right|_{NS5_2} \nn \ee
The Fayet-Iliopoulos (FI) parameters are  determined by the positions of the NS5-branes in the $X^9$ direction, 
\be \zeta = \left.X^9\right|_{NS5_2} - \left.X^9\right|_{NS5_1}\ \ \ ,\ \ \ \tilde{\zeta} = \left.X^9\right|_{NS5_3}-\left. X^9\right|_{NS5_2}\label{fi} \ee

\para
When both FI-parameters are turned on, the vortex states are introduced by adding D1-strings, stretched in the $X^9$ direction, suspended between the NS5-branes and the D3-brane. This is shown in the right-hand side of Figure \ref{fig1}. We introduce $n$ D1-strings attached to the second NS5-brane and a further $\tilde{n}$ attached to the third NS5-brane. Notice that these latter strings have mass $X^9|_{NS5_3}-X^9|_{NS5_1}= \zeta +\tilde{\zeta}$. Comparing this to \eqn{nn}, we see that $n$ and $\tilde{n}$ are identified with the winding of the scalar fields, rather than the flux of the magnetic fields.

\begin{figure}[htb]
\begin{center}
\epsfxsize=3.4in\leavevmode\epsfbox{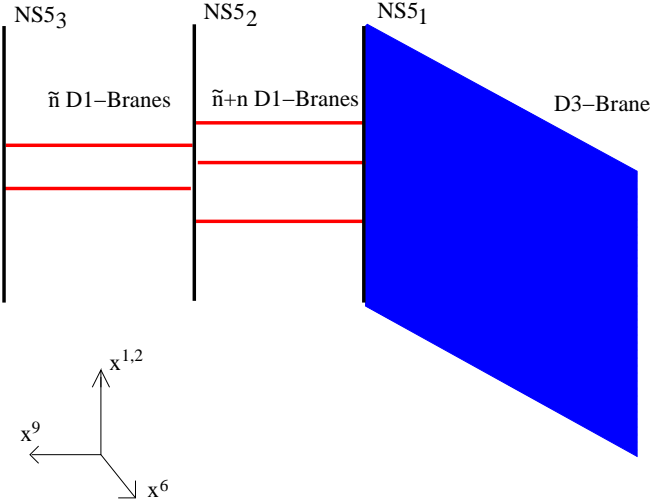}
\end{center}
\caption{The vortex theory on the D1-branes.}
\label{fig2}
\end{figure}

%
%
\para
Our goal now is to identify the theory living on the D1-strings. The quickest way is to momentarily place all NS5-branes to lie at the same position in the $X^6$ direction. The resulting configuration is shown in Figure \ref{fig2}. From this, it is simple to read off the vortex theory on the D1-strings. It is ${\cal N}=(2,2)$ Yang-Mills quantum mechanics with gauge group  $U(\tilde{n})\times U(n+\tilde{n})$. The matter content consists of the following chiral multiplets,
%
%
\be Z_1 &:& \mbox{in the adjoint of $U(\tilde{n})$}\nn\\
Z_2 &:& \mbox{in the adjoint of $U(n+\tilde{n})$}\nn\\
\upsilon\  &:& \mbox{in the bi-fundamental of $U(\tilde{n})\times U(n+\tilde{n})$}\nn\\
\tilde{\upsilon} \ &:& \mbox{in the anti-bi-fundamental of $U(\tilde{n})\times U(n+\tilde{n})$}\nn\\
\varphi \ &:& \mbox{in the fundamental of $U(n+\tilde{n})$}
\nn\ee

There is also a superpotential
\be
{\cal W} = \tilde{\upsilon}(Z_1-Z_2)\upsilon \nn
\ee
with group indices contracted appropriately, which ensures that, in the absence of $\varphi$, the theory has ${\cal N}=(4,4)$ supersymmetry: $\upsilon$ and $\tilde\upsilon$ combine to form a hypermultiplet and $Z_1$ and $Z_2$ combine with the two ${\cal N}=(2,2)$ vector multiplets to form a pair of ${\cal N}=(4,4)$ vector multiplets. The chiral multiplet $\varphi$, which breaks the ${\cal N} = (4.4)$ supersymmetry to ${\cal N}=(2,2)$, arises from the strings stretched between the D1-strings and the D3-brane.

\para
The positions of the NS5-branes determine the parameters of the D1-brane theory which can therefore be related to the parameters of the original $d=2+1$ dimensional theory.  The gauge coupling constants of $U(\tilde{n})$ and $U(n+\tilde{n})$ are given by
\be\frac{1}{g_1^2}\sim \tilde{\zeta} \ \ \ ,\ \ \ \frac{1}{g_2^2}\sim\zeta\label{gz}\ee
%
%
respectively\footnote{This equation hides a sin. The  suppressed factors of $g_s$ and $\alpha'$ are different on the two sides (as, indeed, they have to be on dimensional grounds). This has consequence. It means that the D1-brane theory and the D3-brane theory are valid in different regimes of string theory parameters. This is not uncommon when describing solitons in terms of D-branes (e.g. instantons and monopoles). However, in the present case it means that, while the Higgs branch of the D1-brane theory coincides with the vortex moduli space, the metrics do not agree: they are ``renormalised" as one interpolates between the two regimes \cite{vib}.}. Meanwhile, moving the NS5-branes back to their rightful positions in the $X^6$ direction induces FI parameters in the D1-brane theory. These are 
\be r_1\sim \frac{1}{\tilde{e}^2}\ \ \ ,\ \ \ r_2\sim \frac{1}{e^2}\nn\ee
%
This provides a relationship between the K\"ahler class of the vortex moduli space and the gauge coupling constants in $d=2+1$ dimensions. 

\para
The vortex moduli space is identified with the Higgs branch of the D1-brane theory. The F-term conditions arising from $Z_1$ and $Z_2$ set $\tilde{\upsilon}=0$. Meanwhile, the vector multiplets give rise to a pair of D-term conditions,
\be [Z_1,Z_1^\dagger] + \upsilon\upsilon^\dagger = r_1\ \ \ ,\ \ \ \ [Z_2,Z_2^\dagger] - \upsilon^\dagger\upsilon + \varphi\varphi^\dagger = r_2\label{dterms}\ee
Each of these equations transforms in the adjoint of one factor of the gauge group (with the other group indices on $\upsilon$ summed over appropriately). Finally, the F-term condition for $\tilde{\upsilon}$ sets
\be Z_1 \upsilon - \upsilon Z_2 = 0 \label{lastf}\ee
while the F-term condition for $v$ is trivially satisfied in view of the fact that $\tilde v = 0$.

\para
The Higgs branch is defined by the constraints \eqn{dterms} and \eqn{lastf}, modulo $U(\tilde{n})\times U(n+\tilde{n})$ gauge transformations. It is a $2(n+\tilde{n})$ dimensional manifold. Roughly speaking, the eigenvalues of $Z_1$ can be thought of as the position of the $\tilde{n}$ vortices. The requirement \eqn{lastf} will typically then fix $\tilde{n}$ eigenvalues of $Z_2$; the remaining eigenvalues can be thought of as the positions of the remaining $n$ vortices. 

\para
The Higgs branch naturally inherits a K\"ahler metric from the quotient construction above.  As we mentioned in the footnote, this metric does not generally coincide with the metric on the vortex moduli space. Nonetheless, if one is interested in BPS information, protected by supersymmetry, then the Higgs branch description provides a useful substitute for the full vortex dynamics.

\subsubsection*{Freezing Vortices}

The Higgs branch above describes the dynamics of the $n$ vortices of mass $\zeta$ and the $\tilde{n}$ vortices of mass $\zeta+\tilde{\zeta}$. As we saw in the previous section, freezing the latter results in  $n$ vortices moving in the presence of magnetic impurities. 

\para
 It is simple to implement this freezing in the D1-brane theory. As we see from \eqn{fi} and \eqn{gz}, the limit $\tilde{\zeta}\rightarrow \infty$ corresponds to sending  $X^9\vert_{NS5_3} \rightarrow \infty$ and $g_1^2\rightarrow 0$, so that the third NS5-brane recedes into the distance and the $U(\tilde{n})$ factor of the gauge group becomes a global symmetry. This results in the impurity D-brane model introduced in \cite{shamit}. However, viewing this model as a limit of the product gauge group allows us to understand what background D-brane fields survive.  Specifically, the adjoint field $Z_1$ has kinetic term $\Tr\, |\dot{Z}_1|^2/g_1^2$ and ceases to propagate. Nonetheless, it continues to take a non-zero value, obeying \eqn{dterms} and \eqn{lastf} and now has an interpretation as the distribution of  impurities. 
 
 \para
 This is perhaps best illustrated with a simple example. We describe a single vortex moving in the presence of a single background  delta-function impurity. From our discussion above, the relevant theory on D1-branes is $U(1)\times U(2)$ quantum mechanics. $Z_1$ is now a neutral complex scalar which can be thought of as the position of the impurity. $Z_2$ is a complex $2\times 2$ matrix transforming in the adjoint of $U(2)$; one of its eigenvalues is fixed by $Z_1$ and the other eigenvalue can be thought of as the position of the vortex. The complex scalar $\upsilon$ has charge $+1$ under $U(1)$ and transforms in the $\bf{\bar{2}}$ of $U(2)$.

\para
The first D-term condition \eqn{dterms} reads
 \be \sum_{a=1}^2\upsilon^\dagger_a\upsilon_a = r_1\nn\ee
 Using the $U(2)$ gauge symmetry, we are free to set $\upsilon_2=0$ and take $\upsilon_1$ to be real, so $\upsilon_1= \sqrt{r_1}$. This choice breaks $U(1)\times U(2)\rightarrow U(1)\times U(1)$.  

\para
 We now turn to the F-term condition \eqn{lastf}. This can be written as
\be  \left(\begin{array}{cc} z' & w \\ w' & z\end{array}\right) \left(\begin{array}{c}v_1 \\ 0\end{array}\right) - \left(\begin{array}{c}v_1 \\ 0\end{array}\right) z_1 = \left(\begin{array}{c} 0 \\ 0\end{array}\right)\nn\ee
which sets $z'=z_1$ and $w'=0$.

\para
Finally, we're left with the $U(2)$ D-term condition. This results in two real constraints and a single complex constraint. The real constraints are
\be |\varphi_1|^2 + |w|^2 &=& r_1+r_2\nn\\
|\varphi_2|^2 -|w|^2 &=&r_2  \nn\ee
These can be thought of as moment-map conditions for the surviving $U(1)\times U(1)$ gauge symmetry, under which $\varphi_1$ has charge $(+1,0)$, $\varphi_2$ has charge $(0,+1)$ and $w$ has charge $(+1,-1)$. This defines a complex manifold of dimension 1.

\para
Finally, the complex parameters $z$ and $z_1$ encode the positions of the vortex and the impurity. They are given in terms of the parameters $\phi_1$, $\phi_2$ and $w$ by complex constraint,
\be (z-z_1)w^\dagger + \varphi_1^\dagger\varphi_2=0\nn\ee
This provides an algebraic geometric description of the moduli space of a single vortex in the background of an inserted impurity.

\section*{Acknowledgements}

We thank Nick Dorey, Shamit Kachru, Nick Manton and Gonzalo Torroba for useful discussions and comments. We are supported by STFC and by the European Research Council under the European Union's Seventh Framework Programme (FP7/2007-2013), ERC Grant agreement STG 279943, Strongly Coupled Systems

\end{document}